\documentclass[acmlarge,nonacm]{acmart}

\usepackage{subcaption}
\usepackage{url} 
\AtBeginDocument{%
  }

\begin{document}

\title[Personalized Driver State Modeling Using Physiological Signals in Real-World Automated Driving]{Human-Centered Non-Intrusive Driver State Modeling Using Personalized Physiological Signals in Real-World Automated Driving}

\author{David Puertas-Ramirez}
\email{dpuertasr@dia.uned.es}
\affiliation{%
  \institution{UNED}
  \city{Madrid}
  \country{Spain}
}

\author{Raul Fernandez-Matellan}
\email{raulfern@ing.uc3m.es}
\affiliation{%
  \institution{UC3M}
  \city{Madrid}
  \country{Spain}
}

\author{David Martin Gomez}
\email{dmgomez@ing.uc3m.es}
\affiliation{%
  \institution{UC3M}
  \city{Madrid}
  \country{Spain}
}

\author{Jesus G. Boticario}
\email{jgb@dia.uned.es}
\affiliation{%
  \institution{UNED}
  \city{Madrid}
  \country{Spain}
}

%\authorrunning{Puertas-Ramirez et al.}
\renewcommand{\shortauthors}{Puertas-Ramirez et al.}

\begin{abstract}
    In vehicles with partial or conditional driving automation (SAE Levels 2-3), the driver remains responsible for supervising the system and responding to take-over requests. Therefore, reliable driver monitoring is essential for safe human–automation collaboration. However, most existing Driver Monitoring Systems rely on generalized models that ignore individual physiological variability. In this study, we examine the feasibility of personalized driver state modeling using non-intrusive physiological sensing during real-world automated driving. We conducted experiments in an SAE Level 2 vehicle using an Empatica E4 wearable sensor to capture multimodal physiological signals, including electrodermal activity, heart rate, temperature, and motion data. To leverage deep learning architectures designed for images, we transformed the physiological signals into two-dimensional representations and processed them using a multimodal architecture based on pre-trained ResNet50 feature extractors. Experiments across four drivers demonstrate substantial interindividual variability in physiological patterns related to driver awareness. Personalized models achieved an average accuracy of 92.68\%, whereas generalized models trained on multiple users dropped to an accuracy of 54\%, revealing substantial limitations in cross-user generalization. These results underscore the necessity of adaptive, personalized driver monitoring systems for future automated vehicles and imply that autonomous systems should adapt to each driver's unique physiological profile.
\end{abstract}




%\received{19 January 2026}

\maketitle

\section{Introduction}

The division of driving tasks and responsibilities is evolving as vehicles become more automated. According to SAE International \cite{SAE2021}, Level 0 vehicles are fully manual, while Level 1 is termed Driver Assistance. Level 2, or Partial Driving Automation, provides sustained control of both steering and speed, yet requires the driver to continuously supervise the system and the driving environment. At Level 3, vehicles can handle most driving tasks, but occasionally issue a Take-Over Request (TOR) to the driver. Level 4 systems operate autonomously within a restricted Operational Design Domain (ODD), and Level 5 represents a vision of complete autonomy in all driving conditions.

Current implementations consist of vehicles with Partial Driving Automation (SAE Level 2), and Level 4 "RoboTaxis" deployments limited in geo-fenced areas \cite{NHTSA-Accidents}. This focuses attention on the complexities of the intermediate levels where driving tasks demand collaboration \cite{SAE2021}. A fundamental challenge for both Level 2 and Level 3 is the well-known paradox of the "irony of automation" \cite{Bainbridge1983}, which states that the more reliable a system is, the harder it is for the human operator to maintain attention. At Level 2, this applies to the task of continuous supervision. At Level 3, it applies to the driver's ability to maintain a "fallback-ready" state for TORs, a dynamic that also introduces complex legal and ethical questions about determining responsibility at any given moment \cite{euaiact2024}. Resolving the challenges of both sustained supervision and safe control transitions fundamentally depends on accurately understanding the driver's state.

Existing Driver Monitoring Systems (DMS) often rely on user-agnostic models to assess attention or alertness \cite{Capallera2023}, overlooking significant inter-individual variability in physiological signals, cognitive load, and behavior, especially under complex driving conditions \cite{wang2024}. This generalization can compromise accuracy, particularly for diverse drivers. 

Current research has begun exploring the physiological state of users inside Autonomous Vehicles (AVs) using simulation environments, as this way the complete environment can be controlled and dangerous situations can be tested safely \cite{Rosique2023, Beckers2023, Deng2024, Arakawa2019}. While simulations provide valuable insights and a controlled environment for testing, they have limitations and may not fully capture the intricacies, challenges, and unexpected situations inherent in real-world data, which are essential for gaining a deeper understanding of the problem \cite{Ahlstrom2018, Lotz2019, Reagan2019}.

Our work addresses these two distinct gaps: we propose a novel approach that not only uses physiological sensors in a real-world driving environment but also develops personalized models to accurately reflect each individual's state. To guide our investigation, we formulate the following hypotheses:

\begin{itemize}
    \item \textbf{{H1}:} A personalized modeling approach, using data from non-intrusive physiological sensors in real-world driving, can accurately classify distinct driver states.
    
    \item \textbf{{H2}:} The physiological patterns of driver awareness exhibit significant and measurable inter-individual variability.
    
    \item \textbf{{H3}:} Incorporating data from other drivers into a user-specific training set degrades the model's predictive accuracy compared to a model trained exclusively on that user's own data.

\end{itemize}

The experiments for this study were conducted using an experimental SAE Level 2 vehicle in real-world driving conditions on public highways. To model the driver's state, we used data from non-intrusive physiological sensors. This real-world methodology ensures that participant responses are natural and immersion is high, enabling the identification of unexpected situations and potential problems relevant to future commercial deployments. The core of our modeling technique utilizes Convolutional Neural Networks (CNNs), which are well-established for complex pattern recognition. Since CNNs are designed for image data, we first transform the raw physiological signals into 2D image representations suitable for this type of analysis.

The value of personalization is well-established in fields like precision medicine and adaptive learning, where user-specific models outperform generalized ones \cite{saha2020, diedrichsen2010, Gil2019, Klatzky2023, Tran2021, FCollins2015, Brusilovsky2007}. Adopting this approach shifts the experimental focus from broad data collection to intensive, longitudinal data from fewer participants to effectively model individual differences.

\paragraph{Contributions.}
This work investigates the feasibility and implications of personalized driver-state modeling using non-intrusive physiological sensing in real-world automated driving. The main contributions of this paper are:

\begin{enumerate}
\item \textbf{Real-world physiological sensing for automated driving.} \\
We present a driver monitoring study conducted in a real-world SAE Level 2 automated vehicle, using non-intrusive wearable sensors to capture multimodal physiological signals under natural driving conditions.

\item \textbf{A signal-to-image modeling pipeline for physiological data.} \\
We propose a multimodal architecture that converts physiological time-series signals into image representations and leverages pre-trained convolutional neural networks (\textit{ResNet50}) for feature extraction and driver state classification.

\item \textbf{Empirical evidence of strong inter-individual variability in driver awareness.} \\
Through extensive experiments across four drivers, we demonstrate that physiological patterns related to driver awareness vary significantly between individuals, leading to substantial performance degradation when models are applied across users.

\item \textbf{Evidence supporting personalized driver monitoring systems.} \\
Our results show that personalized models dramatically outperform generalized approaches (92.68\% vs.\ 54\% accuracy), suggesting that adaptive personalization is a key design requirement for future human-centered driver monitoring systems in automated vehicles.
\end{enumerate}

The remainder of the paper is organized into the following sections: \ref{RelatedWork} Related Work, \ref{Methodology} Methodology, \ref{Results} Experimental results, \ref{Discussion} Discussion, and \ref{Conclusion} Conclusions and Future Work.

\section{Related Work} \label{RelatedWork}

\subsection*{Driver State Monitoring in Conditional Automation}
The necessity of maintaining situational awareness in SAE Level 2 and 3 vehicles is well-documented \cite{Bainbridge1983, Collet2019}. Research has largely bifurcated into detecting physiological alertness such as drowsiness \cite{Kim2022}, fatigue \cite{Li2023}, and distraction \cite{Yang2019, Bonyani2021}, and modeling the driver's relational dynamics with the vehicle, including trust \cite{Hunter2022, Huang2022}, acceptance \cite{Zhang2023}, and cognitive workload \cite{Chen2023}. While these studies highlight the complexity of the "Fallback Ready State" \cite{Puertas-Ramirez-HAAPIE21}, they often grapple with the challenge of individual variability in defining these states.

\subsection*{Experimental Environments and Sensing Modalities}
The majority of studies rely on simulation to ensure safety and controllability \cite{Rosique2023, Beckers2023, Deng2024}. However, simulations often fail to replicate the stochastic nature and psychological pressure of real-world driving \cite{Ahlstrom2018, Lotz2019, Puertas-Ramirez-ESM2023}. Regarding instrumentation, sensor selection is largely dictated by these environmental constraints. While more intrusive modalities (e.g., Electroencephalogram, Electromyogram) remain prevalent in laboratory settings due to their high signal fidelity \cite{Arakawa2019, Giorgi2023}, real-world constraints necessitate non-intrusive alternatives. Consequently, these studies favor multisensorial approaches combining Heart Rate Variability and Electrodermal Activity to balance data reliability with user comfort and safety \cite{Lohani2019, Tavaloski2021, Nacpil2021}.

\subsection*{Methodological Trends: 2D Encoding and Personalization}
To leverage the feature extraction capabilities of Convolutional Neural Networks (CNNs), recent efforts have moved beyond raw 1D signal processing. A growing body of work utilizes techniques such as Markov Transition Fields \cite{Yan2022RollingNetwork}, Gramian Angular Fields \cite{Xu2020HumanNetwork}, and Recurrence Plots \cite{Marwan2007} to encode physiological time-series into 2D images \cite{Elalamy2021Multi-modalsignals, Lee2021DrivingSignals}. Furthermore, the field is witnessing a pivot from inter-subject (generalized) to intra-subject (personalized) modeling. Evidence from related domains, such as education and affective computing, suggests that accounting for individual baselines significantly enhances model sensitivity compared to population-based approaches \cite{bobrovitz1998comparison, hayashibe2017personalized, Cabestrero2018SomeStudents, Serrano-Mamolar2021}.

\section{Methodology} \label{Methodology}
This section details the methodology used in our study. We first describe the experimental setup and the signal-to-image transformation techniques. We then provide details on the study participants and the ethical considerations, and conclude by presenting the architecture of our proposed classification model.

\subsection{Experimental Set up} 
    
   The experimental platform utilized in this study has been previously characterized in \cite{Puertas-Ramirez-ESM2023}. It consists of a Toyota Prius PHEV equipped with Comma OpenPilot software \cite{openpilot}, corresponding to an SAE Level 2 vehicle. In this configuration, both longitudinal (acceleration and braking) and lateral (steering) control are managed by the intelligence system, while the driver remains responsible for oversight and must be prepared to take over if necessary.
    
    In the OpenPilot system, user modeling is based solely on a front camera that detects whether the driver is looking at the road. This approach lacks personalization and has multiple failure points. To address these shortcomings, this methodology incorporates physiological signals, with a focus on non-intrusive devices that are easy to wear and do not disturb the driver. This investigation aims to see whether we can extract meaningful results from such devices in comparison to more accurate, invasive alternatives. In fact, for real-world applications, it has been found that sensors for Electrodermal Activity (EDA) and Heart Rate (HR) provide higher applicability than more invasive Electromyography (EMG) and Electroencephalogram (EEG) devices \cite{Lohani2019}. Therefore, we utilized an Empatica E4 wristband, a non-intrusive device that collects seven distinct data streams: EDA, HR, Skin Temperature (TEMP), Blood Volume Pulse (BVP), and accelerometer data along the x, y, and z axes (ACC{x}, ACC{y}, ACC{z}).
    
\begin{figure}[tbh]
        \centering
        \includegraphics[width=\linewidth]{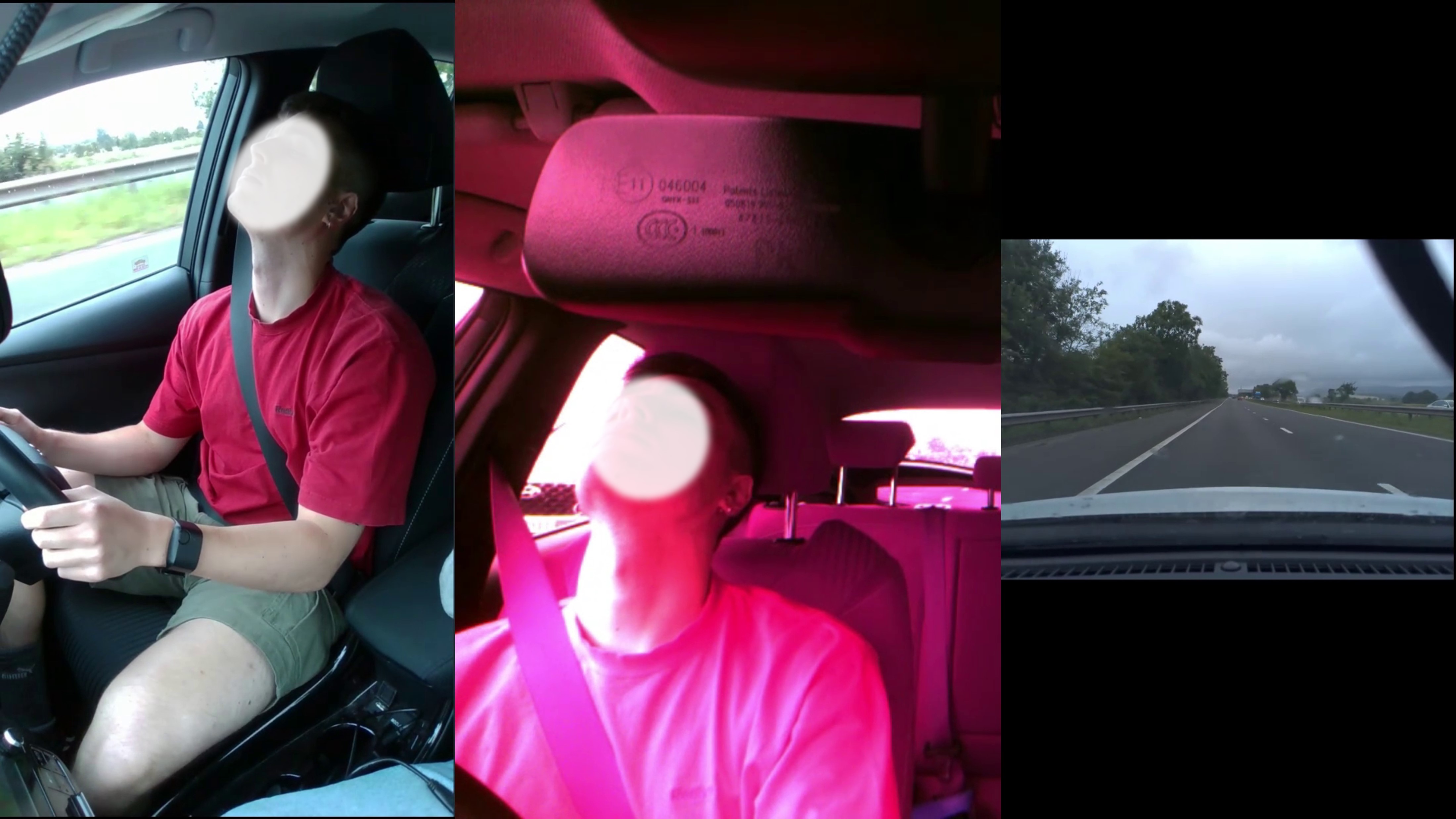}
        \caption{Snapshot of human centered DMS recording in a Real World Scenario. From left to right: Intel RealSense (Color), OpenPilot (Interior), OpenPilot(Exterior)}
        \Description{Composition of three images during an experiment where the user is looking up while holding the steering wheel: (left) color camera showing whole body, (middle) infrared camera showing upper body, (right) camera showing the road in front of the vehicle}
        \label{fig:RealExperiment-merge}
    \end{figure}

    All experiments were carried out under real-world driving conditions, allowing the vehicle to interact autonomously with other road users and facilitating the capture of more realistic affective driver states than would be possible in a simulator. Figure \ref{fig:RealExperiment-merge} shows a snapshot of the real driving environment during one of the experiments.

    Following the methodology detailed in \cite{Puertas-Ramirez-ESM2023}, each experimental session was structured around 16 distinct scenarios, designed to elicit a range of driver states. These scenarios consisted of different permutations of several conditions, including: lane changes, head position (side, up, or down), looking at the road (yes/no), hands on the steering wheel (yes/no), periods of induced inattention, and the occurrence of an unexpected event. To ensure consistency, reproducibility, and safety, a human co-pilot continuously monitored the driving environment and enacted systematically each scenario. The experiments were conducted on public highways and were paused if the co-pilot detected any potentially dangerous road conditions.

\subsection{Physiological Signal to Image transformation}

    Empatica E4 wristband provides the following signals at their respective sampling frequencies BVP at 64 Hz, ACC{x}, ACC{y} and ACC{z} at 32 Hz, EDA and TEMP at 4 Hz and HR at 1 Hz. These signals are one-dimensional, as each time stamp corresponds to a single value for the variable being measured. The measurement time plays a crucial role in this context, as past states influence future ones, especially in affective computing. Detecting trends, such as peaks and troughs (e.g. local maxima and minima), is essential for identifying physiological changes over time.

    To capture these temporal dynamics, we need methods that can effectively process time intervals. Instead of processing sequences directly, we transform the time intervals of the signals into matrices, which are then converted into images. This approach allows the use of pre-trained feature extractors from robust image-based neural networks trained on large datasets. In addition, the use of matrices allows us to exploit the full potential of 2D convolutional neural networks (CNNs), while preparing the model for future integration with facial or body image data.
    
    \subsubsection{Signal Preprocessing}

    The methods for converting one-dimensional signals into two-dimensional images are designed to work regardless of the number of data points in the signal, which is determined by the product of the sampling frequency and the window size. Note that the generated images are squares of size \(sample\_frequency \times window\_size\). To simplify the process and reduce computational complexity, we chose to resample all signals to a common frequency of 4Hz. This decision was based not only on previous studies in the educational domain \cite{Serrano-Mamolar2021}, which indicate the importance of EDA in inferring affective states, but also on ensuring that the signals remain manageable for our methodology without excessive manipulation. Resampling at higher frequencies, such as 32 Hz, would result in much larger images (e.g., a 30-second window would result in 960x960 pixel images), which could increase computation time, limiting real-time applications.

    \subsubsection{Important parameters while generating images from signals}
    
\begin{figure}[tbh]
                \centering
                \includegraphics[width=1\linewidth]{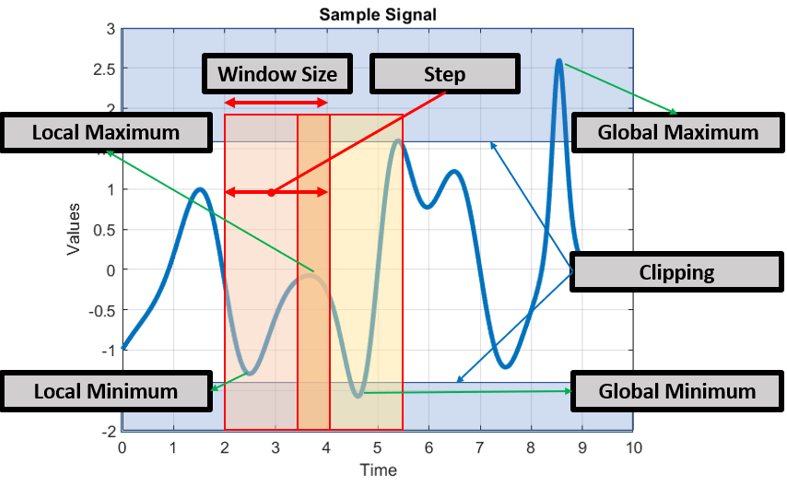}
                \caption{Visual representation of parameters during image generation from physiological signals: local/global maxima and minima, saturation values, working window width and step between consecutive images. }
                \label{fig: Image_Generation}
                \Description{A sample graph illustrating an example where the global and local maximum and minimum peaks are highlighted along with their respective local values. The temporal context of the working window is also shown, with a highlighted section representing the window size and step. A clipping area is displayed, indicating the boundaries of the signal domain. This visualization illustrates how to analyze signal behavior over time.}
    \end{figure}

    In the pre-processing step, all signals were resampled to 4Hz. As all signals are on the same time axis (from the Empatica E4 sensor), they are synchronized by default. The following are the key parameters that influence the image generation process and are illustrated in Figure \ref{fig: Image_Generation}.

     \begin{itemize}
        \item \textbf{Window Size:} Defines the time interval for each generated image, set to 30 seconds in the experiments.
        \item \textbf{Step:} Represents the time shift between successive images, experimentally set to 3 seconds.
        \item \textbf{Scaling:} Ensures image values are in the range [0, 255] using equation (\ref{ecuacion: Scaling signal}):
        
        \begin{equation}
            \label{ecuacion: Scaling signal}
            \overline{x}_{scale}^i= 255 \frac{x^i-\min_{x}}{\max_{x}-\min{x}}
        \end{equation}
        
        Three normalization methods were considered:
        \begin{itemize}
            \item \textbf{Global Metrics:} Uses the entire signal for normalization, preserving scale but reducing contrast.
            \item \textbf{Local Metrics:} Uses only the sliding window, enhancing contrast but losing relative scale.
            \item \textbf{Combined Metrics:} Balances global and local normalization with parameter \(\epsilon=0.5\), formulated in equation (\ref{ecuacion: Normalisation signal}).
            
            \begin{equation}
                \label{ecuacion: Normalisation signal}
                \overline{x}_{end}^i=\epsilon(\overline{x}_{global-metrics}^i) +(1-\epsilon)(\overline{x}_{local-metrics}^i)
            \end{equation}
        \end{itemize}
        
        \item \textbf{Clipping:} To mitigate noise and extreme values, signals are clipped at the 95\% percentile (set to 255) and 5\% percentile (set to 0).
    \end{itemize}

         All of these parameters were fine-tuned manually on the basis of observations of how they affected the quality of the generated images. After initial evaluation, the Combined Metrics normalization method was selected for all subsequent experiments. This choice, along with the other parameters, was kept consistent across all comparisons to isolate the influence of different image generation methods. The following sections explain different techniques for converting a signal into an image.
    
    \subsubsection{Recurrence Plot (RP)}
    
    It is a tool used to visualize the repetitive patterns of a trajectory \(\Vec{x} \in \mathbb{R}^{d}\) in the phase space \cite{Marwan2007RecurrenceSystems}. The method is adapted to one-dimensional signals \(d=1\), making the term-by-term differences.  Two configurations are considered: (1) a binary RP using a fixed threshold (\(\epsilon=0.5\), chosen empirically after testing several values) and the Heaviside function  \(\Theta\) for binarization, and (2) a continuous RP formed by omitting both the threshold and the Heaviside function, resulting in a richer representation of the signal dynamics.

    \subsubsection{Graminian Angular field (GAF)}

    GAF transforms a one-dimensional time series into a 2D image by first normalizing the data (between [0,1]) and then mapping it to polar coordinates \cite{Yan2022RollingNetwork}. The angles are used to construct either the Gramian Angular Summation Field (GASF) or the Gramian Angular Difference Field (GADF) via trigonometric functions of the sum or difference of these angles. These images capture temporal correlations and patterns of the original series.

    \subsubsection{Markov Transition Field (MTF)}
    MTF generates a state-transition matrix by defining discrete states in the time series and computing first-order transition probabilities, resulting in a transition matrix \(W\). The matrix is then mapped back to the time indices of the original series to form a two-dimensional representation, where each element indicates the probability of moving from one state to another over time \cite{Xu2020HumanNetwork}. Two configurations were analyzed: one with 4 states and another with 128 states, each providing a different level of granularity in capturing the probabilistic dynamics of the series.

 \subsection{Participants}
    The study involved four male volunteers capable of operating a Toyota Prius SAE Level 2 vehicle. The participant pool was limited to certified operators, as driving the experimental vehicle on real roads required specific university permission and inclusion on the car's insurance. All participants had previous experience with Comma's OpenPilot software, ensuring a high level of safety during real-world autonomous driving. Since the primary goal was to model individual driver behavior, each participant participated in an intensive session rather than pooling data from many users. Three participants (User 1, User 2, and User 4) were between the ages of 25 and 29, while one participant (User 3) was between the ages of 45 and 49.
    
    \subsection{Ethical and Legal Concerns}

    All ethical and legal considerations were integrated throughout the entire methodological design to ensure compliance with institutional and regulatory standards. Participants provided informed consent following ethical principles and institutional guidelines. Data collection, processing, and storage adhered to strict protocols in full compliance with European Union legislation on data protection and cybersecurity \cite{ENISA2024DataProtection}, safeguarding participants' rights and privacy. See the "Ethical Statement"  section for further details.

    \subsection{Dataset Characteristics}

We analyzed four distinct datasets, which were collected from four male participants (N = 4). Three of the participants were between the ages of 25 and 29, and one was between the ages of 45 and 49. The datasets contained 2,158, 1,288, 825, and 754 images, respectively. Labeling was performed according to established literature to maintain consistency and remove potential bias \cite{Saneiro2014}. Each image was then categorized into one of four awareness levels: Low-Low (LL), Low (L), High (H), and High-High (HH). This 2-bit state representation was chosen to provide greater granularity than binary models, enabling the investigation of a "Fallback Readiness" of the user \cite{SAE2021}. The naming convention adheres to the ISA-18.2.5 alarm management standard \cite{isa_tr18_2_5_2022} and the HL7 medical code system observation standard \cite{hl7_standard}. As detailed below, the class distribution for these levels is unique to each user.

    \begin{itemize}
        \item User 1: 429 LL, 776 L, 513 H, 440 HH
        \item User 2: 259 LL, 263 L, 368 H, 398 HH
        \item User 3: 192 LL, 263 L, 133 H, 237 HH
        \item User 4: 137 LL, 246 L, 202 H, 169 HH
        
    \end{itemize}
    
    Images can be generated using any signal-to-image conversion method. For the comparison of different image generation methods shown in the results section, we used User 1's dataset as it has the most amount of images. For this user, we generated six different datasets, varying the image generation techniques while keeping the class distribution and the number of samples per class constant.

    The datasets were divided into three sets: 70\% for training, 15\% for validation, and 15\% for testing. Samples were randomly selected from the entire dataset and assigned to the appropriate training, validation or test set.

    An important aspect of the dataset architecture is its multimodal nature, with images from different signals remaining separate. Each class folder (LL, L, H, HH) contains seven subfolders, each corresponding to a different physiological signal (Acc{x}, Acc{y}, Acc{z}, Temp, EDA, BVP, HR). The images in these subfolders are time-synchronized, and we ensured timestamp alignment by assigning identical filenames to images corresponding to the same time point across signals (e.g. '0001.jpg').

\subsection{Proposed architecture}

\begin{figure*}[tb]
            \centering
            \includegraphics[width=0.8\linewidth]{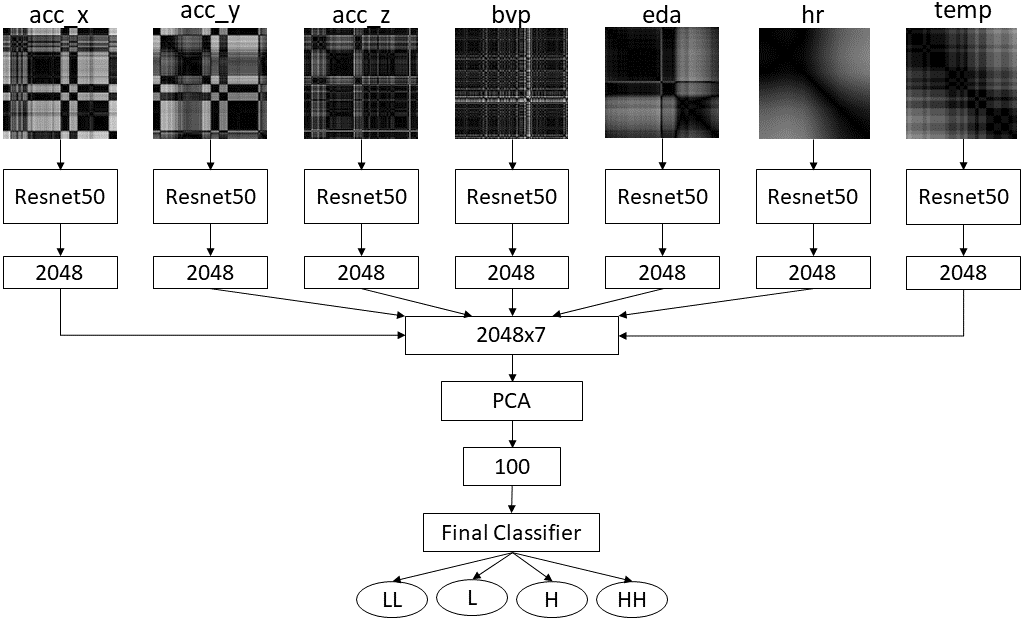}
            \caption{Architecture of the proposed ResNet model}
            \label{fig: Proposed resnet model}
            \Description{Seven inputs (acc_x, acc_y, acc_z, bvp, eda, hr and temp) each goes to a different Resnet50 system. Each of them produces a 2048 wide result. This 2048x7 output is fed into a PCA to reduce it to 100 features. A final classifier gives the final results in one of the levels: LL, L, H, HH}
        \end{figure*}

        We propose a multimodal architecture designed to process each signal independently before fusing the extracted information from all signals. The model takes as input seven images, each generated from the physiological signals recorded by the Empatica E4 sensor: Acc{x}, Acc{y}, Acc{z}, Temp, EDA, BVP and HR. Our approach builds on the architecture developed by \cite{Elalamy2021Multi-modalsignals}, but with important differences. Instead of binary classification, our model classifies into four states (LL,L,H,HH), and instead of training CNN feature extractors from scratch, we use pre-trained ResNet50 using transfer learning. The model architecture is shown in Figure \ref{fig: Proposed resnet model}.
        
        \subsubsection{Feature extraction}
        The first step is to extract features from each of the seven images. Here we apply transfer learning using ResNet50 \cite{He_2016_CVPR}, pre-trained on a large dataset. We extract information from the penultimate layer of ResNet50, compressing each image into a feature vector of size \(2048\times1\). This process is repeated for all seven input images (one per signal), resulting in seven feature vectors. These feature vectors are then concatenated into a single vector of size \(14336\times1\) (because \(2048\times7=14336\). Importantly, no parameters of ResNet50 are trained during this process. 
        
        \subsubsection{Dimensionality Reduction}
        
         After feature extraction, the next step is dimensionality reduction. To reduce the high-dimensional concatenated vector, we applied Principal Component Analysis (PCA). The number of components was selected empirically to find a balance between model complexity and training efficiency. We observed that with 50 components, the model struggled to converge within 25 epochs. Conversely, with 200 components, the model converged very quickly (in 3 epochs), suggesting a higher risk of overfitting. A choice of 100 components provided a good trade-off, with the model converging in a reasonable 13 epochs. Therefore, we reduced the dimensionality from 14,336 features to 100 for all experiments
         
        \subsubsection{Classifier}
        The final stage of the model is a fully connected neural network for classification. This network consists of two fully connected (FC) layers:
            \begin{enumerate}
                \item The first FC layer reduces the 100 input features to 25 using ReLU as the activation function.
                \item The second FC layer further reduces the 25 features to 10, again using ReLU activation.
                \item The output layer then maps the 10 features to 4 classes (LL, L, H, HH) using a softmax activation function. The class with the highest softmax score is selected as the predicted output.
            \end{enumerate}

        \subsubsection{Training Process}
        
        Training was performed on an Nvidia RTX A4000 GPU (16 GB memory) paired with an Intel Core i9 CPU using the PyTorch framework. The model was trained with the following parameters Loss function: Cross-Entropy Loss, Optimizer: Adam Learning Rate: 0.001 Weight Decay: \(e^-4\).
        We saved the model with the highest validation F1-score during training. To ensure statistical robustness, the training process was repeated five times for each configuration. The final results reported in the Results section represent the mean performance and standard deviation across these five independent runs on each user's test sets.

\section{Results} \label{Results}

This section details the experimental evaluation of the proposed driver awareness monitoring framework. The analysis is structured into three key phases: first, an evaluation of personalized models to determine the performance ceiling; second, an analysis of the combined-user approach to quantify generalization limits; and finally, a direct comparative analysis to validate the statistical significance of the findings.

\subsection{Comparative of signal-to-images methods}

To determine the most effective feature extraction technique, several state-of-the-art signal-to-image transformation methods were evaluated. These experiments were conducted using the dataset from User 1, which provided the highest number of images, thus ensuring a robust comparison. Key parameters, such as a 30-second window size and a 3-second step between consecutive images, were kept constant across all methods. For this experiment, we determined the best performance using validation accuracy.

As detailed in Table \ref{tab:encoding_comparison}, the encoding method significantly influenced classification performance. Continuous Recurrence Plots (RP) yielded the highest accuracy at 70.33\%, outperforming all other techniques. In contrast, Binary RP showed a significant drop in performance, with an accuracy of 61.9\%. The GASF achieved an accuracy of 60.11\%, while the GADF had the lowest accuracy of 49.70\%. The MTF provided 63.39\% accuracy with 4 levels, but dropped to 58.9\% when 128 levels were used. Overall, Continuous RP consistently provided the best results and was selected as the preferred method for converting signals into images for the remainder of the study.

    \begin{table}[hbt!]
    \caption{Impact of Signal-to-Image Encoding Methods on Accuracy (User 1). Continuous Recurrence Plots (RP) provided the most robust feature representation.}
    \label{tab:encoding_comparison}
    \centering
    \begin{tabular}{lc}
    \toprule
    \textbf{Encoding Method} & \textbf{Accuracy} \\
    \midrule
    \textbf{Continuous RP} & \textbf{70.33\%} \\
    MTF (4 levels) & 63.39\% \\
    Binary RP & 61.90\% \\
    GASF & 60.11\% \\
    MTF (128 levels) & 58.90\% \\
    GADF & 49.70\% \\
    \bottomrule
    \end{tabular}
    \end{table}

\subsection{Personalized Model Performance and Inter-Subject Variability} \label{subsec:personalized_performance}

To evaluate the efficacy of subject-specific training (Hypothesis H1) and the extent of physiological variability between drivers (Hypothesis H2), we trained a personalized model for each of the four users. To ensure statistical reliability, the training represents the mean and standard deviation across all five independent runs\footnote{The complete results dataset and statistical computations are provided as supplementary material.}.

Strict data separation was maintained throughout the process, using distinct training, validation, and testing sets to prevent overfitting. Regularization techniques, including weight decay, dropout, and L1 regularization, were also explored but did not result in a significant improvement in performance.

\textbf{Performance of Personalized Models (H1):}
As shown by the diagonal values (bold) in Table \ref{Table: Personalized training}, the personalized models achieved high accuracy across all subjects. The accuracy ranged from 87.52\% (User 1) to 96.52\% (User 4), with low standard deviations between $\pm 0.6\%  $ and $\pm 1.6\%$. These results suggest that a personalized modeling approach can accurately classify distinct driver awareness states using non-intrusive sensor data.

\textbf{Assessment of Inter-Subject Variability (H2):}
To measure how well these features transfer between people, each personalized model was evaluated on the other three users (off-diagonal values). A significant decrease in performance was observed in all cross-subject tests. While models achieved $>$90\% accuracy on the user they were trained on, accuracy dropped to between 40.62\% and 63.52\% when applied to a different user. 

For example, the model trained on User 2 achieved 91.79\% accuracy on User 2 but only 50.22\% when tested on User 1. This difference indicates that physiological patterns vary significantly between individuals and are not easily transferable. These findings support Hypothesis H2, showing that individual variability is a major challenge for generalized modeling.

        \begin{table*}[hbt!]
        \caption{Cross-Subject Accuracy Matrix. Rows represent the user being tested, and columns represent the user data used for training. The diagonal (bold) highlights the personalized performance (mean $\pm$ standard deviation).}
        \label{tab:cross_user_accuracy}
        \centering
        \begin{tabular}{lcccc}
        \toprule
        & \multicolumn{4}{c}{\textbf{Trained on}} \\
        \cmidrule(lr){2-5}
        \textbf{Tested on} & \textbf{User 1} & \textbf{User 2} & \textbf{User 3} & \textbf{User 4} \\
        \midrule
        \textbf{User 1} & \textbf{87.52 $\pm$ 0.78\%} & 50.22 $\pm$ 2.29\% & 58.37 $\pm$ 1.95\% & 51.85 $\pm$ 3.31\% \\
        \textbf{User 2} & 46.26 $\pm$ 2.60\% & \textbf{91.79 $\pm$ 0.63\%} & 41.64 $\pm$ 3.58\% & 40.62 $\pm$ 4.12\% \\
        \textbf{User 3} & 63.52 $\pm$ 3.47\% & 50.88 $\pm$ 2.23\% & \textbf{94.88 $\pm$ 1.21\%} & 57.44 $\pm$ 1.54\% \\
        \textbf{User 4} & 61.39 $\pm$ 4.83\% & 48.00 $\pm$ 1.78\% & 56.17 $\pm$ 2.58\% & \textbf{96.52 $\pm$ 1.63\%} \\
        \bottomrule
        \end{tabular}
        \label{Table: Personalized training}
        \end{table*}

\subsection{Evaluating a Combined-User Modeling Approach}

    \label{subsec:combined_user_modeling}

    To analyze the effects of aggregating physiological data from multiple subjects and to test the feasibility of a generalized system (Hypothesis H3), a combined-user approach was evaluated. To ensure robust validation and prevent potential data leakage between training and test sets, a random sampling validation scheme was employed at the user level.

    Specifically, four independent models were generated, each trained on a group of three subjects. To assess both generalization capability and the impact of data aggregation, every model was subsequently evaluated against the individual test sets of all four users. This allows for a dual analysis: (a) performance on the unseen user (excluded from training), and (b) performance on the seen users (included in training). As with the personalized experiments, every training configuration was repeated five times to ensure statistical significance.
    
    The results, presented in Table \ref{Table: Combined training}, highlight two critical challenges associated with the combined-data approach.
    
    \textbf{Generalization to Unseen Users:}
    First, the models demonstrated limited ability to generalize to new individuals. As shown by the diagonal elements in Table \ref{Table: Combined training} (where the test user was excluded from the training mix), accuracy ranged from 45.23\% to 57.55\% across the four awareness states. This represents a substantial drop compared to the $>$87\% accuracy achieved by personalized models. For instance, the model trained on Users 1, 3, and 4 achieved only 45.23\% accuracy when tested on the unseen User 2. While this study prioritizes a high-density subject-specific data collection strategy, the results obtained under these constraints suggest that physiological signatures are highly individual-specific. Consequently, these findings indicate that a generalist model relying on physiological signals may struggle to capture the universal patterns necessary to accurately classify an unseen driver without further adaptation.

    \textbf{Impact of Data Mixing:}
    Second, even when a user's data was included in the training set (off-diagonal elements), a consistent performance penalty was observed compared to the purely personalized baselines. By comparing the off-diagonal values in Table \ref{Table: Combined training} with the personalized results in Table \ref{Table: Personalized training}, drops in accuracy ranging from roughly 5\% to 12\% are evident. 
    
    For example, User 3 achieved 94.88\% accuracy with a dedicated model, but this dropped to 87.04\% when their data was combined with Users 1 and 4. This indicates that physiological patterns are highly subject-specific. Consequently, including data from other individuals appears to introduce conflicting feature representations that hinder the model's ability to accurately classify the target user's awareness state.

        \begin{table*}[hbt!]
        \caption{Combined-User Accuracy Matrix. Columns represent the specific subset of users used for training. The diagonal (bold) highlights the ''Leave-One-Out'' performance, where the tested user was excluded from the training set.}
        \label{tab:combined_user_accuracy}
        \centering
        \begin{tabular}{lcccc}
        \toprule
        & \multicolumn{4}{c}{\textbf{Trained on Users}} \\
        \cmidrule(lr){2-5}
        \textbf{Tested on} & \textbf{\{2, 3, 4\}} & \textbf{\{1, 3, 4\}} & \textbf{\{1, 2, 4\}} & \textbf{\{1, 2, 3\}} \\
        \midrule
        \textbf{User 1} & \textbf{57.55 $\pm$ 3.22\%} & 82.26 $\pm$ 1.64\% & 83.57 $\pm$ 1.43\% & 81.88 $\pm$ 0.87\% \\
        \textbf{User 2} & 86.26 $\pm$ 1.51\% & \textbf{45.23 $\pm$ 2.91\%} & 85.23 $\pm$ 2.62\% & 82.56 $\pm$ 4.51\% \\
        \textbf{User 3} & 91.36 $\pm$ 1.43\% & 87.04 $\pm$ 3.01\% & \textbf{57.12 $\pm$ 3.03\%} & 85.76 $\pm$ 3.01\% \\
        \textbf{User 4} & 86.61 $\pm$ 1.80\% & 85.57 $\pm$ 3.82\% & 86.09 $\pm$ 1.23\% & \textbf{55.13 $\pm$ 2.99\%} \\
        \bottomrule
        \end{tabular}
        \label{Table: Combined training}
        \end{table*}

\subsection{Comparative Analysis: Personalized vs. Combined}

To synthesize the findings and compare the efficacy of the two modeling strategies, a direct statistical comparison was performed between the subject-specific models (Section \ref{subsec:personalized_performance}) and the combined-user models (Section \ref{subsec:combined_user_modeling}). Table \ref{tab:overall_results} summarizes the performance metrics across the entire dataset.

\textbf{Overall Performance Gap:}
The personalized approach demonstrated significantly higher performance metrics compared to the combined approach within this study. As detailed in Table \ref{tab:overall_results}, the personalized models achieved an average classification accuracy of \textbf{92.68\%}, whereas the combined models averaged 53.76\%—a difference of nearly 40 percentage points. Statistical analysis (paired t-test) indicates that these differences are significant ($p < 0.001$) for Accuracy, Precision, Recall, and F1-Score. These results suggest that, for the specific physiological features analyzed, a personalized approach yields a substantially more robust classification than a combined one.

\textbf{Class-wise Sensitivity and Inconsistency:}
To investigate the source of the performance gap, the recall (sensitivity) for each awareness state was analyzed individually. Recall was selected as the primary metric for this comparison to assess the system's ability to correctly identify each specific state, as failing to detect a state of low awareness carries a significantly higher risk in a safety-critical context. While other metrics exhibit similar performance trends, recall provides the most direct insight into the reliability of the monitoring system (full metrics available in the supplementary material). Figure \ref{fig:class_wise_comparison} contrasts the recall of the Personalized models (Blue) against the Combined models (Orange) for all four users.

The analysis reveals a high degree of variability and inconsistency in the combined models' performance across different subjects. This instability is further quantified by the standard deviations of the recall metrics: while personalized models exhibited stable performance with narrow deviations (typically $<5\%$), the combined models demonstrated significantly larger dispersion, reaching up to 15\% for specific states (e.g., User 3, `High' state). 

Consequently, the combined models exhibited erratic behavior. For instance, the combined model achieved high recall for User 3's `Low-Low' state (96\%), yet performed poorly on the same state for User 2 (49\%). Similarly, while it captured User 1's `Low' state reasonably well (76\%), it missed the majority of `High' state instances for Users 2 and 4 (26\% and 36\%, respectively).

This lack of consistency presents a challenge for a combined/generalized approach. The results indicate that these combined models do not merely perform worse on average, but fail in non-uniform ways depending on the specific physiological profile of the driver. Consequently, these findings imply that a combined model may struggle to provide a consistent safety baseline, as its predictive capability appears to fluctuate unpredictably between users.

\begin{table*}[hbt!]
\caption{Overall Performance Comparison. The Personalized approach significantly outperforms the Combined (Generalized) approach across all metrics ($p < 0.001$). Values are reported as Mean $\pm$ Standard Deviation. Bold values indicate the best performance in each column.}
\label{tab:overall_results}
\begin{tabular}{lcccc}
\toprule
\textbf{Model Approach} & \textbf{Accuracy} & \textbf{Precision} & \textbf{Recall} & \textbf{F1-Score} \\
\midrule
Personalized & \textbf{92.68 $\pm$ 3.67\%} & \textbf{92.88 $\pm$ 3.49\%} & \textbf{92.64 $\pm$ 3.65\%} & \textbf{92.67 $\pm$ 3.57\%} \\
Combined (General) & 54.00 $\pm$ 6.00\% & 55.28 $\pm$ 3.31\% & 53.93 $\pm$ 5.40\% & 52.76 $\pm$ 4.87\% \\
\midrule
\textit{Significance ($p$)} & \textit{$< 0.001$} & \textit{$< 0.001$} & \textit{$< 0.001$} & \textit{$< 0.001$} \\
\bottomrule
\end{tabular}
\label{Table: Global Results}
\end{table*}

\begin{figure*}[hbt!]
    \centering
    \includegraphics[width=\textwidth]{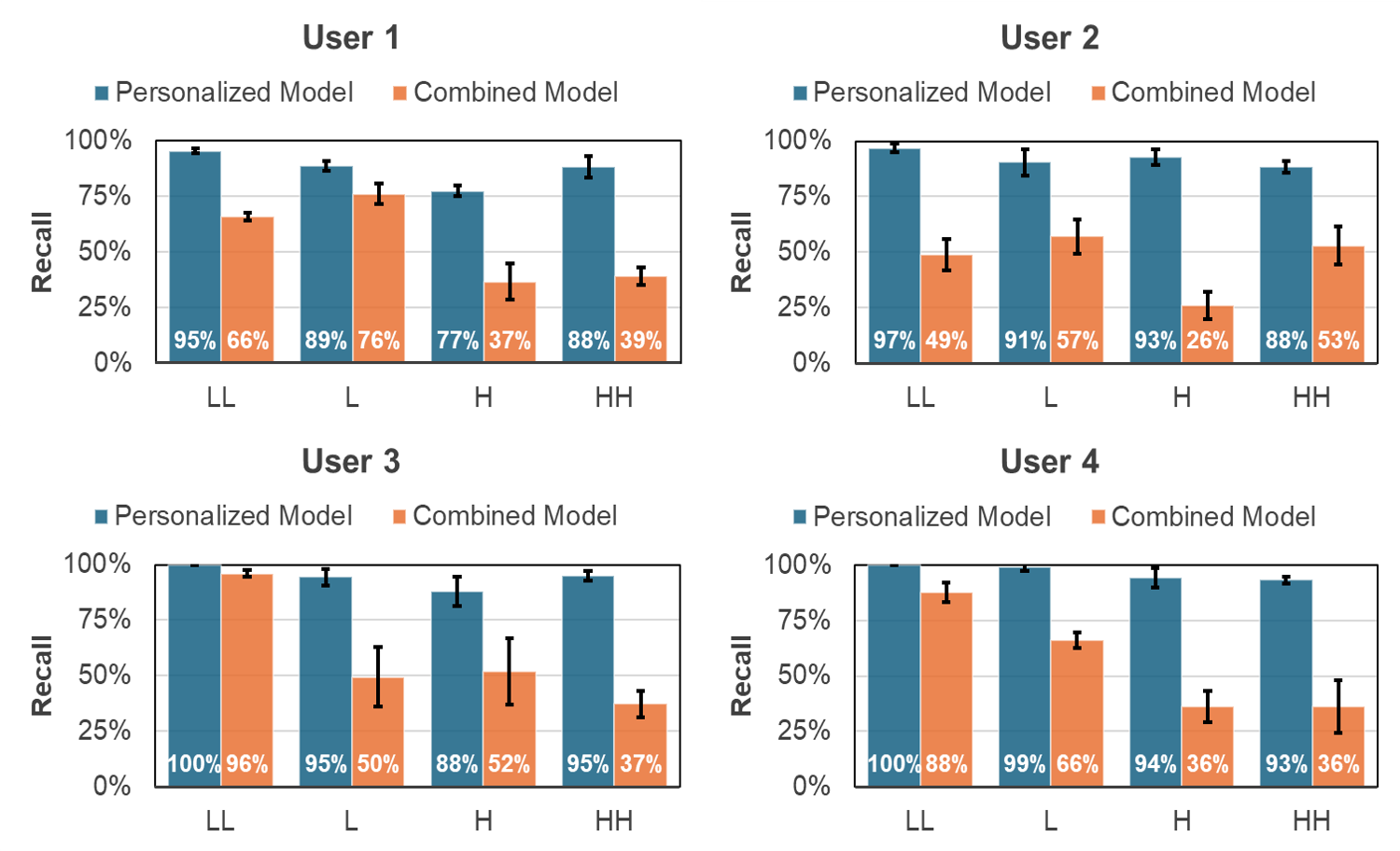}
    \caption{Class-wise Recall Analysis. The Personalized models (Blue) maintain high sensitivity across all awareness states. In contrast, the Combined models (Orange) show significant degradation.}
    \Description{A matrix of four bar charts, one for each study participant (Users 1–4). Each chart compares the recall of a Personalized Model (blue bars) against a Combined Model (orange bars) across four awareness states: Low-Low, Low, High, and High-High. The personalized models maintain consistently high recall above 90\%, while the combined models show high variability and significant performance drops}
    \label{fig:class_wise_comparison}
\end{figure*}

\section{Discussion} \label{Discussion}

In vehicles with partial or conditional driving automation (SAE Levels 2-3), the driver remains integral to the driving task, responsible for supervision and, when necessary, intervention. This critical human-in-the-loop role necessitates reliable methods for assessing the driver's state. This study addresses this challenge by evaluating the viability of a personalized approach, with results providing strong support for the three foundational hypotheses. While prior work has successfully used physiological sensors in controlled simulators \cite{Rosique2023, Beckers2023}, this study validates the approach within the unpredictability of real-world driving. This provides crucial evidence that non-intrusive sensors can capture meaningful data despite the noise and variability inherent in on-road conditions, a critical step toward deployment.

\textbf{Validation of Personalized Modeling (H1):}
In support of the first hypothesis (H1), the results demonstrate that personalized models can accurately classify a driver's state, achieving an average accuracy of \textbf{92.68\%}. The signal-to-image conversion technique proves fundamental, effectively enabling the application of robust, pre-trained CNN architectures like ResNet50 to physiological time-series data. Demonstrating the transferability of such well-established computer vision models to this domain is of significant interest to the scientific community, suggesting that off-the-shelf deep learning architectures can be effectively leveraged for high-precision physiological monitoring \cite{Lee2021DrivingSignals}.

\textbf{The Limits of Generalization (H2 \& H3):}
The results also clarify why personalization is so effective. As predicted in H2, significant variability was found between individuals, with accuracy dropping to $\approx$53.76\% in cross-user tests. Supporting H3, mixing data from different drivers consistently degraded performance compared to the personalized baselines. Beyond the average drop in accuracy, the comparative analysis revealed that generalized models exhibit significant instability, with large standard deviations and inconsistent recall across critical awareness states. This aligns with findings in other domains where personalized models have shown significant performance gains over generalized counterparts \cite{li2024comparison, Serrano-Mamolar2021}.

\textbf{Implications for Human-Centered Autonomous Systems:}
This study's results have several implications for designing future driver monitoring systems for automated vehicles.
First, the significant discrepancy in performance between personalized and generalized models indicates that driver-state monitoring must consider individual physiological variability. Systems that rely exclusively on population-level models may fail to provide reliable estimates of driver awareness, which could reduce the effectiveness of safety-critical interventions.
Second, the findings support the development of adaptive driver monitoring architectures that evolve over time. One practical approach would be to deploy an initial generalized model that provides baseline monitoring while gradually collecting data from the driver to train a personalized model. Such hybrid systems would combine immediate availability with long-term accuracy.
Third, personalization may improve human–automation collaboration by enabling more context-aware interactions. For instance, adaptive driver models could facilitate more precise timing of takeover requests, minimize false alarms in attention monitoring, and enable interfaces that adapt to the driver's cognitive state.
Taken together, these results suggest that future driver monitoring systems should transition from static, population-level models to human-centered, adaptive systems that progressively learn each driver's physiological profile.

\textbf{Limitations and Future Work:}
It is important to acknowledge the scope of this study. To ensure the highest data quality and safety during on-road testing, participation was restricted to certified operators. Consequently, the dataset prioritizes data depth over participant count, featuring high-density recordings for a small number of subjects rather than sparse data across a large population. While this strategy was necessary to validate the personalization hypothesis (H1), further research with a larger, more diverse participant pool is needed to explore broader generalization strategies. Additionally, the results are based on a specific set of driving scenarios; expanding the environmental conditions would further test model robustness. Finally, the choice of medical-grade wearable sensors represents a trade-off between real-world practicality and signal resolution \cite{Lohani2019}. These limitations define the scope of the study and underscore that while the proposed hybrid framework is a logical next step, its viability must be confirmed through future research.

\section{Conclusion} \label{Conclusion}

This study evaluated the efficacy of personalized versus generalized modeling for monitoring driver awareness states using non-intrusive physiological sensors in real-world driving conditions. By employing a signal-to-image encoding technique and a ResNet50 architecture, the personalized approach achieved high classification accuracy, averaging 92.68\% across four awareness states. These results provide strong evidence that subject-specific modeling can effectively overcome the noise inherent in on-road environments.

In contrast, the analysis of combined-user models highlighted significant challenges regarding cross-subject generalization. Models trained on multiple users failed to achieve consistent performance on unseen subjects, with accuracies dropping to approximately 53.76\% and exhibiting high variability across different drivers. These findings suggest that physiological signatures of awareness are highly individual-specific, making the implementation of a reliable subject-independent model struggle under the conditions analyzed.

Based on these findings, this study highlights a critical trade-off between the immediate availability of generalized models and the superior accuracy of personalized ones. While a subject-independent system offers the advantage of immediate deployment, our results suggest that it may lack the reliability required for safety-critical driver monitoring due to the high inter-individual variability observed in physiological responses. Consequently, these results suggest that further research into adaptive or transfer learning frameworks is necessary. Such systems could potentially bridge this gap by using generalized parameters as a baseline and progressively adapting to the specific physiological profile of the driver as data becomes available. Future work will focus on evaluating these adaptive strategies across larger and more diverse subject groups to determine the minimum data requirements for effective personalization.

\appendix

\section*{Ethical Statement} \label{Ethical}

The \textbf{European General Data Protection Regulation (GDPR)} \cite{eugdpr2016} has been applied at all stages of this research. The system was designed from the ground up with strict adherence to these regulations, as well as the recommendations and guidelines of our institution. All collected data is registered under our university’s data protection framework.

\section*{Reproducibility}

The dataset utilized in this study contains sensitive personal information, including high-resolution video and physiological recordings. Given the subject-specific nature of the modeling approach, complete anonymization of the raw signal data is not feasible without compromising data integrity; consequently, the full raw dataset cannot be made publicly available. 

However, to ensure scientific reproducibility and transparency:
\begin{itemize}
    \item \textbf{Methodology:} The experimental setup and dataset generation protocol follow the procedures detailed in \cite{Puertas-Ramirez-ESM2023}.
    \item \textbf{Code:} The repository of source code—including the signal-to-image conversion tools, model architectures, specific hyperparameters, and random seeds will be made publicly available upon acceptance of the manuscript.
    \item \textbf{Statistical Data:} An spreadsheet file containing the complete set of performance metrics, including per-run results for all 5 repetitions, means, standard deviations, and statistical test outputs, is provided as supplementary material.
\end{itemize}

\bibliographystyle{ACM-Reference-Format}
\bibliography{base, manual}

@article{FCollins2015,
    title = {{A New Initiative on Precision Medicine}},
    year = {2015},
    journal = {New England Journal of Medicine},
    author = {Collins, Francis S. and Varmus, Harold},
    number = {9},
    month = {2},
    pages = {793--795},
    volume = {372},
    publisher = {New England Journal of Medicine (NEJM/MMS)},
    url = {https://www.nejm.org/doi/full/10.1056/NEJMp1500523},
    doi = {10.1056/NEJMP1500523/SUPPL{\_}FILE/NEJMP1500523{\_}DISCLOSURES.PDF},
    issn = {0028-4793},
    pmid = {25635347}
}

@article{Lohani2019,
    title = {{A Review of Psychophysiological Measures to Assess Cognitive States in Real-World Driving}},
    year = {2019},
    journal = {Frontiers in Human Neuroscience},
    author = {Lohani, Monika and Payne, Brennan R. and Strayer, David L.},
    number = {March},
    month = {3},
    pages = {1--27},
    volume = {13},
    url = {https://www.frontiersin.org/article/10.3389/fnhum.2019.00057/full},
    doi = {10.3389/fnhum.2019.00057},
    issn = {1662-5161}
}

@article{Nacpil2021,
    title = {{Application of Physiological Sensors for Personalization in Semi-Autonomous Driving: A Review}},
    year = {2021},
    journal = {IEEE Sensors Journal},
    author = {Nacpil, Edric John Cruz and Wang, Zheng and Nakano, Kimihiko},
    number = {18},
    month = {9},
    pages = {19662--19674},
    volume = {21},
    publisher = {Institute of Electrical and Electronics Engineers Inc.},
    doi = {10.1109/JSEN.2021.3100038},
    issn = {15581748}
}

@article{Collet2019,
    title = {{Associating vehicles automation with drivers functional state assessment systems: A challenge for road safety in the future}},
    year = {2019},
    journal = {Frontiers in Human Neuroscience},
    author = {Collet, Christian and Musicant, Oren},
    month = {2},
    pages = {408476},
    volume = {13},
    publisher = {Frontiers Media S.A.},
    doi = {10.3389/FNHUM.2019.00131/BIBTEX},
    issn = {16625161}
}

@article{Rosique2023,
    title = {{Autonomous Vehicle Dataset with Real Multi-Driver Scenes and Biometric Data}},
    year = {2023},
    journal = {Sensors},
    author = {Rosique, Francisca and Navarro, Pedro J and Miller, Leanne and Salas, Eduardo},
    number = {4},
    month = {2},
    pages = {2009},
    volume = {23},
    url = {https://www.mdpi.com/1424-8220/23/4/2009},
    doi = {10.3390/s23042009},
    issn = {1424-8220}
}

@article{Chen2023,
    title = {{Comparing driver reaction and mental workload of visual and auditory take-over request from perspective of driver characteristics and eye-tracking metrics}},
    year = {2023},
    journal = {Transportation research part F: traffic psychology and behaviour},
    author = {Chen, Weiya and Sawaragi, Tetsuo and Hiraoka, Toshihiro},
    month = {8},
    pages = {396--410},
    volume = {97},
    publisher = {Pergamon},
    doi = {10.1016/J.TRF.2023.07.012},
    issn = {1369-8478}
}

@inproceedings{He_2016_CVPR,
    title = {{Deep Residual Learning for Image Recognition}},
    year = {2016},
    booktitle = {2016 IEEE Conference on Computer Vision and Pattern Recognition (CVPR)},
    author = {He, Kaiming and Zhang, Xiangyu and Ren, Shaoqing and Sun, Jian},
    month = {6},
    pages = {770--778},
    volume = {2016-Decem},
    publisher = {IEEE},
    url = {http://ieeexplore.ieee.org/document/7780459/},
    address = {Las Vegas, NV, USA},
    isbn = {978-1-4673-8851-1},
    doi = {10.1109/CVPR.2016.90},
    issn = {10636919},
    arxivId = {1512.03385}
}

@article{Gil2019,
    title = {{Designing human-in-the-loop autonomous Cyber-Physical Systems}},
    year = {2019},
    journal = {International Journal of Human-Computer Studies},
    author = {Gil, Miriam and Albert, Manoli and Fons, Joan and Pelechano, Vicente},
    month = {10},
    pages = {21--39},
    volume = {130},
    publisher = {Academic Press},
    doi = {10.1016/J.IJHCS.2019.04.006},
    issn = {1071-5819}
}

@article{Bonyani2021,
    title = {{Dipnet: Driver Intention Prediction for a Safe Takeover Transition in Automated Vehicles}},
    year = {2021},
    journal = {SSRN Electronic Journal},
    author = {Bonyani, Mahdi and Rahmanian, Mina and Jahangard, Simindokht and Rezaei, Mahdi},
    number = {April},
    month = {12},
    publisher = {Elsevier BV},
    url = {https://papers.ssrn.com/abstract=3982515},
    doi = {10.2139/ssrn.3982515},
    issn = {1556-5068}
}

@article{Li2023,
    title = {{Driver fatigue detection and human-machine cooperative decision-making for road scenarios}},
    year = {2023},
    journal = {Multimedia Tools and Applications},
    author = {Li, Anna and Ma, Xinnan and Guo, Jiaxin and Zhang, Jingyue and Wang, Jing and Zhao, Kai and Li, Yaochen},
    month = {7},
    pages = {1--32},
    publisher = {Springer},
    url = {https://link.springer.com/article/10.1007/s11042-023-15994-7 https://link.springer.com/10.1007/s11042-023-15994-7},
    doi = {10.1007/s11042-023-15994-7},
    issn = {1380-7501}
}

@article{Lee2021DrivingSignals,
    title = {{Driving Stress Detection Using Multimodal Convolutional Neural Networks with Nonlinear Representation of Short-Term Physiological Signals}},
    year = {2021},
    journal = {Sensors},
    author = {Lee, Jaewon and Lee, Hyeonjeong and Shin, Miyoung},
    number = {7},
    month = {3},
    pages = {2381},
    volume = {21},
    url = {https://www.mdpi.com/1424-8220/21/7/2381},
    doi = {10.3390/s21072381},
    issn = {1424-8220}
}

@article{Tavaloski2021,
    title = {{HARMONY: A Human-Centered Multimodal Driving Study in the Wild}},
    year = {2021},
    journal = {IEEE Access},
    author = {Tavakoli, Arash and Kumar, Shashwat and Guo, Xiang and Balali, Vahid and Boukhechba, Mehdi and Heydarian, Arsalan},
    pages = {23956--23978},
    volume = {9},
    publisher = {Institute of Electrical and Electronics Engineers Inc.},
    url = {https://ieeexplore.ieee.org/document/9343252},
    doi = {10.1109/ACCESS.2021.3056007},
    issn = {21693536}
}

@article{Zhang2023,
    title = {{Human acceptance of autonomous vehicles: Research status and prospects}},
    year = {2023},
    journal = {International Journal of Industrial Ergonomics},
    author = {Zhang, Qidi and Zhang, Tingru and Ma, Liang},
    pages = {103458},
    volume = {95},
    publisher = {Elsevier},
    url = {https://doi.org/10.1016/j.ergon.2023.103458},
    doi = {10.1016/j.ergon.2023.103458},
    issn = {18728219}
}

@article{Xu2020HumanNetwork,
    title = {{Human Activity Recognition Based on Gramian Angular Field and Deep Convolutional Neural Network}},
    year = {2020},
    journal = {IEEE Access},
    author = {Xu, Hongji and Li, Juan and Yuan, Hui and Liu, Qiang and Fan, Shidi and Li, Tiankuo and Sun, Xiaojie},
    pages = {199393--199405},
    volume = {8},
    url = {https://ieeexplore.ieee.org/document/9234451/},
    doi = {10.1109/ACCESS.2020.3032699},
    issn = {2169-3536}
}

@inproceedings{Puertas-Ramirez-ESM2023,
    title = {{Improving Autonomous Vehicle Automation Through Human-System Interaction}},
    year = {2023},
    booktitle = {The 37th annual European Simulation and Modelling Conference},
    author = {Puertas-Ramirez, David and Fernandez-Matellan, Raul and Martin-Gomez, David and G. Boticario, Jesus and Tena-Gago, David},
    pages = {294--300},
    url = {https://hdl.handle.net/20.500.14468/23191},
    address = {Toulouse, France}
}

@article{saha2020,
    title = {{Intra- and Inter-subject Variability in EEG-Based Sensorimotor Brain Computer Interface: A Review}},
    year = {2020},
    journal = {Frontiers in Computational Neuroscience},
    author = {Saha, Simanto and Baumert, Mathias},
    month = {1},
    pages = {506286},
    volume = {13},
    publisher = {Frontiers Media S.A.},
    url = {www.frontiersin.org},
    doi = {10.3389/FNCOM.2019.00087/BIBTEX},
    issn = {16625188}
}

@article{Bainbridge1983,
    title = {{Ironies of automation}},
    year = {1983},
    journal = {Automatica},
    author = {Bainbridge, Lisanne},
    number = {6},
    month = {11},
    pages = {775--779},
    volume = {19},
    url = {https://linkinghub.elsevier.com/retrieve/pii/0005109883900468},
    doi = {10.1016/0005-1098(83)90046-8},
    issn = {00051098}
}

@article{Beckers2023,
    title = {{JOAN: a framework for human-automated vehicle interaction experiments in a virtual reality driving simulator}},
    year = {2023},
    journal = {Journal of Open Source Software},
    author = {Beckers, Niek and Siebinga, Olger and Giltay, Joris and van der Kraan, André},
    number = {82},
    month = {2},
    pages = {4250},
    volume = {8},
    publisher = {Open Journals},
    url = {https://joss.theoj.org/papers/10.21105/joss.04250},
    doi = {10.21105/joss.04250},
    issn = {2475-9066}
}

@article{Reagan2019,
    title = {{Measuring Adult Drivers’ Use of Level 1 and 2 Driving Automation by Roadway Functional Class}},
    year = {2019},
    journal = {Proceedings of the Human Factors and Ergonomics Society Annual Meeting},
    author = {Reagan, Ian J. and Hu, Wen and Cicchino, Jessica B. and Seppelt, Bobbie and Fridman, Lex and Glazer, Michael},
    number = {1},
    pages = {2093--2097},
    volume = {63},
    isbn = {1071181319631},
    doi = {10.1177/1071181319631225},
    issn = {2169-5067}
}

@inproceedings{Elalamy2021Multi-modalsignals,
    title = {{Multi-modal emotion recognition using recurrence plots and transfer learning on physiological signals}},
    year = {2021},
    booktitle = {2021 9th International Conference on Affective Computing and Intelligent Interaction (ACII)},
    author = {Elalamy, Rayan and Fanourakis, Marios and Chanel, Guillaume},
    month = {9},
    pages = {1--7},
    publisher = {IEEE},
    url = {https://ieeexplore.ieee.org/document/9597442/},
    address = {Nara, Japan},
    isbn = {978-1-6654-0019-0},
    doi = {10.1109/ACII52823.2021.9597442}
}

@article{Giorgi2023,
    title = {{Neurophysiological mental fatigue assessment for developing user-centered Artificial Intelligence as a solution for autonomous driving}},
    year = {2023},
    journal = {Frontiers in Neurorobotics},
    author = {Giorgi, Andrea and Ronca, Vincenzo and Vozzi, Alessia and Aric{\`{o}}, Pietro and Borghini, Gianluca and Capotorto, Rossella and Tamborra, Luca and Simonetti, Ilaria and Sportiello, Simone and Petrelli, Marco and Polidori, Carlo and Varga, Rodrigo and van Gasteren, Marteyn and Barua, Arnab and Ahmed, Mobyen Uddin and Babiloni, Fabio and Di Flumeri, Gianluca},
    volume = {17},
    publisher = {Frontiers Media SA},
    url = {/pmc/articles/PMC10721973/ /pmc/articles/PMC10721973/?report=abstract https://www.ncbi.nlm.nih.gov/pmc/articles/PMC10721973/},
    doi = {10.3389/FNBOT.2023.1240933},
    issn = {16625218},
    pmid = {38107403}
}

@misc{openpilot,
    title = {{openpilot — open source advanced driver assistance system}},
    year = {2024},
    author = {{Comma.ai}},
    url = {https://comma.ai/openpilot}
}

@article{Yang2019,
    title = {{Patterns of Sequential Off-Road Glances Indicate Levels of Distraction in Automated Driving}},
    year = {2019},
    journal = {Proceedings of the Human Factors and Ergonomics Society Annual Meeting},
    author = {Yang, Shiyan and Kuo, Jonny and Lenn{\'{e}}, Michael G.},
    number = {1},
    pages = {2056--2060},
    volume = {63},
    isbn = {1071181319631},
    doi = {10.1177/1071181319631204},
    issn = {2169-5067}
}

@misc{Deng2024,
    title = {{Predicting Driver Takeover Performance in Conditional Automation (Level 3) through Physiological Sensing}},
    year = {2024},
    author = {Min, Deng and Gluck, Aaron and Menassa, Carol and Kamat, Vineet and Li, Da and Brinkley, Julian},
    month = {1},
    url = {http://deepblue.lib.umich.edu/handle/2027.42/191950},
    institution = {Center for Connected and Automated Transportation},
    isbn = {7347647525},
    doi = {10.7302/21951}
}

@article{Klatzky2023,
    title = {{Psychological Science Meets Wearable Cognitive Assistance}},
    year = {2023},
    journal = {Current Directions in Psychological Science},
    author = {Klatzky, Roberta L. and Satyanarayanan, Mahadev},
    number = {6},
    month = {12},
    pages = {446--453},
    volume = {32},
    publisher = {SAGE Publications Inc.},
    url = {https://journals.sagepub.com/doi/10.1177/09637214231187912},
    doi = {10.1177/09637214231187912/ASSET/IMAGES/LARGE/10.1177{\_}09637214231187912-FIG3.JPEG},
    issn = {14678721}
}

@article{Arakawa2019,
    title = {{Psychophysical assessment of a driver's mental state in autonomous vehicles}},
    year = {2019},
    journal = {Transportation Research Part A: Policy and Practice},
    author = {Arakawa, Toshiya and Hibi, Ryosuke and Fujishiro, Taka aki},
    number = {May 2018},
    pages = {587--610},
    volume = {124},
    publisher = {Elsevier},
    url = {https://doi.org/10.1016/j.tra.2018.05.003},
    doi = {10.1016/j.tra.2018.05.003},
    issn = {09658564}
}

@inproceedings{Lotz2019,
    title = {{Recognizing behavioral factors while driving: A real-world multimodal corpus to monitor the driver's affective state}},
    year = {2018},
    booktitle = {LREC 2018 - 11th International Conference on Language Resources and Evaluation},
    author = {Lotz, Alicia and Ihme, Klas and Charnoz, Audrey and Maroudis, Pantelis and Dmitriev, Ivan and Wendemuth, Andreas},
    pages = {1589--1596},
    publisher = {European Language Resources Association (ELRA)},
    url = {https://aclanthology.org/L18-1251/},
    address = {Miyazaki, Japan},
    isbn = {9791095546009}
}

@article{Marwan2007,
    title = {{Recurrence plots for the analysis of complex systems}},
    year = {2007},
    journal = {Physics Reports},
    author = {Marwan, Norbert and Carmen Romano, M. and Thiel, Marco and Kurths, Jürgen},
    number = {5-6},
    month = {1},
    pages = {237--329},
    volume = {438},
    publisher = {North-Holland},
    doi = {10.1016/J.PHYSREP.2006.11.001},
    issn = {0370-1573}
}

@misc{eugdpr2016,
    title = {{Regulation {\{}(EU){\}} 2016/679 of the European Parliament and of the Council of 27 April 2016 on the protection of natural persons with regard to the processing of personal data and on the free movement of such data, and repealing Directive 95/46/EC (General }},
    year = {2016},
    author = {{European Parliament and Council of the European Union}},
    url = {https://eur-lex.europa.eu/eli/reg/2016/679/oj}
}

@article{Yan2022RollingNetwork,
    title = {{Rolling Bearing Fault Diagnosis Based on Markov Transition Field and Residual Network}},
    year = {2022},
    journal = {Sensors},
    author = {Yan, Jialin and Kan, Jiangming and Luo, Haifeng},
    number = {10},
    month = {5},
    pages = {3936},
    volume = {22},
    url = {https://www.mdpi.com/1424-8220/22/10/3936},
    doi = {10.3390/s22103936},
    issn = {1424-8220}
}

@article{Huang2022,
    title = {{Sharing the Road: How Human Drivers Interact with Autonomous Vehicles on Highways}},
    year = {2022},
    journal = {Proceedings of the Human Factors and Ergonomics Society Annual Meeting},
    author = {Huang, Chunxi and Wen, Xiao and He, Dengbo and Jian, Sisi},
    number = {1},
    month = {9},
    pages = {1437--1441},
    volume = {66},
    publisher = {SAGE PublicationsSage CA: Los Angeles, CA},
    url = {http://journals.sagepub.com/doi/10.1177/1071181322661165 https://journals.sagepub.com/doi/10.1177/1071181322661165},
    doi = {10.1177/1071181322661165},
    issn = {2169-5067}
}

@inproceedings{Puertas-Ramirez-HAAPIE21,
    title = {{Should Conditional Self-Driving Cars Consider the State of the Human Inside the Vehicle?}},
    year = {2021},
    booktitle = {Adjunct Proceedings of the 29th ACM Conference on User Modeling, Adaptation and Personalization},
    author = {Puertas-Ramirez, David and Serrano-Mamolar, Ana and Martin Gomez, David and Boticario, Jesus G.},
    month = {6},
    pages = {137--141},
    publisher = {ACM},
    url = {https://dl.acm.org/doi/10.1145/3450614.3462243},
    address = {New York, NY},
    isbn = {9781450383677},
    doi = {10.1145/3450614.3462243}
}

@article{Ahlstrom2018,
    title = {{Stress , fatigue and inattention amongst city bus drivers – an explorative study on real roads within the ADAS {\&} ME project}},
    year = {2018},
    journal = {Paper presented at the 6th International Conference on Driver Distraction and Inattention (DDI2018)},
    author = {Ahlstr{\"{o}}m, Christer and Anund, Anna and Kjellman, Erik Håkansson},
    pages = {1--7},
    volume = {2}
}

@article{Cabestrero2018SomeStudents,
    title = {{Some insights into the impact of affective information when delivering feedback to students}},
    year = {2018},
    journal = {Behaviour and Information Technology},
    author = {Cabestrero, Raúl and Quir{\'{o}}s, Pilar and Santos, Olga C. and Salmeron-Majadas, Sergio and Uria-Rivas, Raul and Boticario, Jesus G and Arnau, David and Arevalillo-Herr{\'{a}}ez, Miguel and Ferri, Francesc J.},
    number = {12},
    pages = {1252--1263},
    volume = {37},
    publisher = {Taylor {\&} Francis},
    url = {https://doi.org/10.1080/0144929X.2018.1499803},
    doi = {10.1080/0144929X.2018.1499803},
    issn = {13623001}
}

@article{Kim2022,
    title = {{Take-Over Requests after Waking in Autonomous Vehicles}},
    year = {2022},
    journal = {Applied Sciences},
    author = {Kim, Won and Jeon, Eunki and Kim, Gwangbin and Yeo, Dohyeon and Kim, Seungjun},
    number = {3},
    month = {1},
    pages = {1438},
    volume = {12},
    url = {https://www.mdpi.com/2076-3417/12/3/1438},
    doi = {10.3390/app12031438},
    issn = {2076-3417}
}

@techreport{SAE2021,
    title = {{Taxonomy and definitions for terms related to driving automation systems for on-road motor vehicles}},
    year = {2021},
    booktitle = {Surface Vehicle Recommended Practice},
    author = {{SAE International}},
    institution = {SAE International},
    doi = {10.4271/J3016{\_}202104}
}

@article{diedrichsen2010,
    title = {{The coordination of movement: optimal feedback control and beyond}},
    year = {2010},
    journal = {Trends in cognitive sciences},
    author = {Diedrichsen, Jörn and Shadmehr, Reza and Ivry, Richard B.},
    number = {1},
    month = {1},
    pages = {31--39},
    volume = {14},
    publisher = {Trends Cogn Sci},
    url = {https://pubmed.ncbi.nlm.nih.gov/20005767/},
    doi = {10.1016/J.TICS.2009.11.004},
    issn = {1879-307X},
    pmid = {20005767}
}

@article{Hunter2022,
    title = {{The Interaction Gap: A Step Toward Understanding Trust in Autonomous Vehicles Between Encounters}},
    year = {2022},
    journal = {Proceedings of the Human Factors and Ergonomics Society Annual Meeting},
    author = {Hunter, Jacob G. and Konishi, Matthew and Jain, Neera and Akash, Kumar and Wu, Xingwei and Misu, Teruhisa and Reid, Tahira},
    number = {1},
    month = {9},
    pages = {147--151},
    volume = {66},
    publisher = {SAGE PublicationsSage CA: Los Angeles, CA},
    url = {http://journals.sagepub.com/doi/10.1177/1071181322661311 https://journals.sagepub.com/doi/10.1177/1071181322661311},
    doi = {10.1177/1071181322661311},
    issn = {2169-5067}
}

@article{Saneiro2014,
    title = {{Towards emotion detection in educational scenarios from facial expressions and body movements through multimodal approaches}},
    year = {2014},
    journal = {Scientific World Journal},
    author = {Saneiro, Mar and Santos, Olga C. and Salmeron-Majadas, Sergio and Boticario, Jesus G.},
    volume = {2014},
    doi = {10.1155/2014/484873},
    issn = {1537744X}
}

@article{Brusilovsky2007,
    title = {{User Models for Adaptive Hypermedia and Adaptive Educational Systems}},
    year = {2007},
    journal = {Lecture Notes in Computer Science (including subseries Lecture Notes in Artificial Intelligence and Lecture Notes in Bioinformatics)},
    author = {Brusilovsky, Peter and Mill{\'{a}}n, Eva},
    pages = {3--53},
    volume = {4321 LNCS},
    publisher = {Springer, Berlin, Heidelberg},
    url = {https://link.springer.com/chapter/10.1007/978-3-540-72079-9_1},
    isbn = {978-3-540-72079-9},
    doi = {10.1007/978-3-540-72079-9{\_}1},
    issn = {1611-3349}
}

@article{Tran2021,
  author = {Tran, T. N. T. and Felfernig, A. and Tintarev, N.},
  title = {Humanized recommender systems: state-of-the-art and research issues},
  journal = {ACM Transactions on Interactive Intelligent Systems},
  year = {2021},
  volume = {11},
  issue = {2},
  pages = {1-41},
  doi = {10.1145/3446906}
}

@article{Serrano-Mamolar2021,
author = {Serrano-Mamolar, Ana and Arevalillo-Herr{\'{a}}ez, Miguel and Chicote-Huete, Guillermo and {G. Boticario}, Jesus G.},
doi = {10.3390/s21051777},
issn = {1424-8220},
journal = {Sensors},
month = {mar},
number = {5},
pages = {1777},
publisher = {Multidisciplinary Digital Publishing Institute},
title = {{An Intra-Subject Approach Based on the Application of HMM to Predict Concentration in Educational Contexts from Nonintrusive Physiological Signals in Real-World Situations}},
url = {https://www.mdpi.com/1424-8220/21/5/1777},
volume = {21},
year = {2021}
}

@techreport{ENISA2024DataProtection,
  author       = {{European Union Agency for Cybersecurity (ENISA)}},
  title        = {Engineering Personal Data Protection in EU Data Spaces},
  year         = {2024},
  institution  = {ENISA},
  url          = {https://www.enisa.europa.eu/publications/engineering-personal-data-protection-in-eu-data-spaces},
  note         = {Accessed: 2025-06-30}
}

@article{Marwan2007RecurrenceSystems,
    title = {{Recurrence plots for the analysis of complex systems}},
    year = {2007},
    journal = {Physics Reports},
    author = {Marwan, Norbert and Carmen Romano, M. and Thiel, Marco and Kurths, Jürgen},
    number = {5-6},
    month = {1},
    pages = {237--329},
    volume = {438},
    publisher = {North-Holland},
    doi = {10.1016/J.PHYSREP.2006.11.001},
    issn = {0370-1573}
}

@misc{hl7_standard,
  author       = {{Health Level Seven International}},
  title        = {Code System ObservationInterpretation},
  howpublished = {Web Page},
  url          = {https://terminology.hl7.org/CodeSystem-v3-ObservationInterpretation.html},
  urldate      = {2025-07-17},
  year         = {2025},
  note         = {Provides definitions for L, H, LL, and HH codes.}
}

@techreport{isa_tr18_2_5_2022,
  author      = {{International Society of Automation}},
  title       = {Alarm System Monitoring, Assessment, and Auditing},
  institution = {International Society of Automation},
  number      = {ISA-TR18.2.5-2022},
  year        = {2022},
  url         = {https://www.isa.org/products/isa-tr18-2-5-2022-alarm-system-monitoring-assessme},
  note        = {Technical report supplementing the main ISA-18.2 standard.}
}

@misc{euaiact2024,
  author={{{European Parliament} and {Council}}},
  title = {{Regulation (EU) 2024/1689 of the European Parliament and of the Council of 13 June 2024 laying down harmonised rules on artificial intelligence and amending Regulations (EC) No 300/2008, (EU) No 167/2013, (EU) No 168/2013, (EU) 2018/858, (EU) 2018/1139 and (EU) 2019/2144 and Directives 2014/90/EU, (EU) 2016/797 and (EU) 2020/1828 (Artificial Intelligence Act)}},
  year = {2024},
  month = {7},
  day = {12},
  publisher = {Official Journal of the European Union},
  note = {L 177/1-137},
  url = {https://eur-lex.europa.eu/legal-content/EN/TXT/?uri=CELEX:32024R1689}
}

@article{Capallera2023,
  author={Capallera, Marine and Angelini, Leonardo and Meteier, Quentin and Khaled, Omar Abou and Mugellini, Elena},
  journal={IEEE Transactions on Intelligent Vehicles}, 
  title={Human-Vehicle Interaction to Support Driver's Situation Awareness in Automated Vehicles: A Systematic Review}, 
  year={2023},
  volume={8},
  number={3},
  pages={2551-2567},
  doi={10.1109/TIV.2022.3200826}}

@article{wang2024,
  title={The association between physiological and eye-tracking metrics and cognitive load in drivers: a meta-analysis},
  author={Wang, A. and Huang, C. and Wang, J. and He, D.},
  journal={Transportation Research Part F: Traffic Psychology and Behaviour},
  volume={104},
  pages={474--487},
  year={2024},
  doi={10.1016/j.trf.2024.06.014},
  url={https://doi.org/10.1016/j.trf.2024.06.014}
}

@misc{NHTSA-Accidents,
  author       = {{National Highway Traffic Safety Administration}},
  title        = {Standing General Order on Crash Reporting},
  year         = {2025},
  url          = {https://www.nhtsa.gov/laws-regulations/standing-general-order-crash-reporting},
  note         = {Accessed: 2025-07-30}
}

@article{li2024comparison,
  title={A comparison of personalized and generalized approaches to emotion recognition using consumer wearable devices: Machine learning study},
  author={Li, Joe and Washington, Peter and others},
  journal={JMIR AI},
  volume={3},
  number={1},
  pages={e52171},
  year={2024},
  publisher={JMIR Publications Inc., Toronto, Canada}
}

@article{bobrovitz1998comparison,
  title={Comparison of visual inspection and statistical analysis of single-subject data in rehabilitation research},
  author={Bobrovitz, Candace D and Ottenbacher, Kenneth J},
  journal={American journal of physical medicine \& rehabilitation},
  volume={77},
  number={2},
  pages={94--102},
  year={1998},
  publisher={LWW}
}

@incollection{hayashibe2017personalized,
  title={Personalized modeling for home-based postural balance rehabilitation},
  author={Hayashibe, Mitsuhiro and Gonz{\'a}lez, Alejandro and Fraisse, Philippe},
  booktitle={Human Modelling for Bio-Inspired Robotics},
  pages={111--137},
  year={2017},
  publisher={Elsevier}
}

\end{document}